\documentstyle[12pt]{article}\topmargin=0.5in\oddsidemargin=5mm
\begin{document}\def\t{\times}\def\p{\phi}
\def\P{\Phi}\def\a{\alpha}\def\e{\varepsilon}\def\be{\begin{equation}}
\def\ee{\end{equation}}\def\l{\label}\def\0{\setcounter{equation}{0}}
\def\b{\beta}\def\S{\Sigma}\def\C{\cite}\def\r{\ref}\def\ba{\begin{eqnarray}}
\def\ea{\end{eqnarray}}\def\n{\nonumber}\def\R{\rho}\def\X{\Xi}\def\x{\xi}
\def\La{\Lambda}\def\la{\lambda}\def\d{\delta}\def\s{\sigma}\def\f{\frac}
\def\D{\Delta}\def\pa{\partial}\def\Th{\Theta}\def\o{\omega}\def\O{\Omega}
\def\th{\theta}\def\ga{\gamma}\def\Ga{\Gamma}\def\h{\hat}\def
\rar{\rightarrow}\def\vp{\varphi}\def\inf{\infty}\def\le{\left}
\def\ri{\right}\def\foot{\footnote}
\begin{center}
{\Large\bf Generating functionals method of N.N.Bogolyubov and
multiple production physics}\\

\vskip 1cm {\it J.Manjavidze\foot{Permanent address: Inst. of Physics,
Tbilisi, Georgia}, A.Sissakian}\\ JINR, Dubna, Russia \end{center}

\vskip 1cm
\begin{abstract}
The generating functionals (GF) method in Bogolyubov's formulation
and its application for particle physics is considered. Effectiveness
of the method is illustrated by two examples. So, GF method can be
used as the technical trick solving the infinite sequence of
algebraic equations. We will consider the example, where GF allows
express the multiplicity distributions (topological cross sections)
through the particles correlation functions (inclusive cross
sections) to `predict' so called the Koba-Nielsen-Olesen scaling. We
will use the GF to define validity of the thermal description of the
multiple production phenomena also. It will be seen that this will
lead to the `correlations relaxation condition' of N.N.Bogolyubov.
This will allow to offer the experimentally measurable criteria of
applicability of thermodynamical description of multiple production
processes.  In results we will find the closed form of perturbation
theory applicable for kinetic phase of nonequilibrium processes. It
is shown a way as the approach may be adapted to the definite
external conditions.

\end{abstract}

\section{Introduction}\0

It is hard to imagine modern particles physics without such
fundamental notions as, for instance, the phase transitions,
topological defects, taken from statistical physics. This extremely
fruitful connection among two branches of physics based on the
euclidean postulate \C{schw}:  the formulae of particle physics are
coincide with corresponding formulae of statistical physics if the
transformation $t\to it$ is applied.  But this coincidence exist iff
the media is equilibrium only, since the time order of physical
process becomes lost after the transition to imaginary time $it$. So,
the particles static properties only can be considered by euclidean
field theories.

The euclidean postulate does not `work' for arbitrary element of
$S$-matrix and, by this reason, there is not, at first glance,
general connection between particles and statistical
physics. Our aim is demonstrate this connection considering the
multiple production example, staying in the real-time theory frame.

The multiple production is a typical dissipative process of the
incident kinetic energies transition into the energies (masses) of
produced particles. This is the nonequilibrium process and the
fluctuations, generally speaking, may be high in it.  Experimental
data confirms this general expectation at the mean multiplicities
region, when $n\sim \bar{n}$ \C{drem}.

Considering multiple production we would like to note firstly that
the mean multiplicity $\bar{n}$ of hadrons for modern accelerator
energies ($\sim$10 Tev) is large $\bar{n}(s)\simeq$100. So, it is
practically impossible to describe the system with $N=3\bar{n}-10
\simeq$ 300 degrees of freedom using ordinary methods.

Secondly, it is natural to assume that the entropy ${\cal S}$ tends to
maximum with rising multiplicity $n$ and reach the maximum at
${n} \sim n_{max}\sim \sqrt{s}$, since the dissipation take place
in the vacuum (presumably with zero energy density)\foot{This
consideration lie in the basis of earliest Fermi-Landau `statistical'
model of hadrons multiple production.}. But the experiment shows that
at high energies $n\sim\bar{n}(s) \sim \ln^2s<< n_{max}(s)$ are
essential.  This means that there is not total dissipation of
incident energy in considered thermalization process \C{sis1}.
Absence of thermalization may be a consequence of hidden conservation
laws \C{zakh}.

We would like to adopt following fundamental principle of
nonequilibrium statistics introduced by N.N.Bogolyubov \C{bog}. It is
natural to assume that the system evaluate to the equilibrium in such
a way that the `nonequilibrium' fluctuations in it should tend to
zero.  In the frame of Bogolyubov's principle the quantitative measure
of `nonequilibrium' fluctuation is the mean value of correlation
functions and, therefore, this quantities should tend to zero when
the media tends to equilibrium.

In our interpretation the Bogolyubov's correlations relaxation
principle means following. So, for nonequilibrium state presence
`nonequilibrium' fluctuations in the form of the macroscopic flow of,
for instance, energy $\e$ is natural. Then the mean value of
$m$-point correlation functions $K_m$ can not be small as the
consequence of macroscopic flow. But in vicinity of equilibrium the
macroscopic flows should relax and, accordingly, the $mean$ value of
correlation functions should be small, $K_m\approx0$.  To
characterize the equilibrium one may consider also the particles,
charge, spin, etc.  densities macroscopic flows and theirs relaxation.

We would like to show in result that the correlations relaxation
principle leads to the quantitative connection with real time
thermodynamics of Schwinger-Keldysh type\foot{Last one includes the
nonequilibrium thermodynamics also.} \C{kel}. Just for this purpose
the generating functionals (GF) method of Bogolyubov will be used
since it allows to find the quantitative connections, where the
euclidean postulate does not applicable.

We will use more natural for particles physics microcanonical
formalism. In this formalism the thermodynamical `rough' variables
are introduced as the Lagrange multipliers of corresponding
conservation laws. Theirs physical meaning are defined by
corresponding equations of state. So, if the fluctuations in vicinity
of solutions of corresponding equations are Gaussian then one can use
this variables for description of the system. Corresponding
condition is the Bogolyubov's correlations relaxation condition.

Formally, the generating functions method presents the integral
transformation to new variables.  One can choose them as the `rough'
thermodynamical variables. To describe the far from equilibrium
system we will introduce the `local equilibrium hypothesis'. In its
frame the preequilibrium state consist from equilibrium domains. In
this case new variables should depend on the coordinates of domain
and, in result, we are forced to use the generating functionals (GF)
formalism.

We will consider two example to illustrate effectiveness of the GF
method. In Sec.2 we will consider the transformation
(multiplicity $n\to$ activity $z$) to show the origin of the
Koba-Nielsen-Olesen scaling (KNO-scaling)\foot{In privet discussion
with one of the authors (A.S.) at summer of 1973 Z.Koba noted that
the main reason of investigation leading to the KNO-scaling was just
the GF method of N.N.Bogolyubov.}.

In Sec.3 we will investigate a possibility of temperature description
of the multiple production processes. We will consider for this
purpose the transformation (particles energies $\e\to$ temperature
$1/\b$) to find the $S$-matrix interpretation of thermodynamics. It
will be shown that this interpretation would be rightful iff the
correlations are relax.

In Sec.4 we will use this interpretation to formulate the
perturbation theory in the case when $\b$ and $z$ are local
coordinates of temperature $(x,t)$ \C{15}. One can use this closed
form of perturbation theory for description of nonequilibrium media
(in kinetic phase) and for description of the multiple production
process as well.

\section{KNO-scaling}\0

We would like start from note that the generating functions method
allows connect inclusive spectra $f_k$ \C{incl} and exclusive cross
sections $\s_n(s)$. One can use for this purpose the normalization
 condition:
\be
\bar{f}_k\s_{tot}\equiv\int
d\o_k(q)f_k(q_1,q_2,...,q_k)
=\sum_{n=k}\f{n!}{(n-k)!}\s_n,~\bar{f}_k\equiv0~{\rm если}~k>n_{max},
\l{1}\ee
where, as usual, $$d\o_k(q)=\prod^{k}_{i=1}{d^3q_i}/
{(2\pi)^32\e(q_i)},~\e(q)=\sqrt{q^2+m^2}$$, is the Lorentz-covariant
element of phase space.

Eq.(\r{1}) can be considered as the set of coupled equations for
$\s_n$. One may multiply both sides of (\r{1}) on $(z-1)^k/k!$ and
sum over $k$ to solve them. We will see that this is equivalent of
introduction of `big partition function' $\Xi(z)$, where $z$ is the
`activity': the chemical potential $\mu\sim\ln z$.

We will find in result of summation over $k$ that
\be
\Xi(z)\equiv\sum_k\f{(z-1)^k}{k!}\bar{f}_k=\sum_n
z^n\f{\s_n}{\s_{tot}}.
\l{4}\ee
Then, assuming that $\Xi(z)$ is known,
\be
\s_n=\s_{tot}\f{1}{2\pi i}\oint_C\f{dz}{z^{n+1}}\Xi(z),
\l{5}\ee
where the closed contour $C$ includes point $z=0$. Here $\Xi(z)$ is
defined by left hand side of (\r{4}) and is the generating function
of $\s_n$.

The coefficients $C_m$ in decomposition:
\be
\ln\Xi(z)=\sum_m\f{(z-1)^m}{m!}C_m.
\l{6}\ee
are the (binomial) correlators. Indeed,
\be
C_1=\bar f_1=\bar{n},~
C_2=\bar f_2-\le\{\bar f_1\ri\}^2,~
C_3=\bar f_3-3\bar f_2\le\{\bar f_1\ri\}^2+
2\le\{\bar f_1\ri\}^3
\l{7}\ee
an so on. If $C_m=0$, $m>1$, then $\s_n$ is described by Poisson
formulae:
\be
\s_n=\s_{tot}e^{-\bar n}\f{(\bar n)^n}{n!}.
\l{8}\ee
It corresponds to the case of absence of correlations.

Let us consider more week assumption:
\be
C_m(s)=\ga_m\le(C_1(s)\ri)^m,
\l{9}\ee
where $\ga_m$ is the energy independent constant. Then
\be
\ln\Xi(z,s)=\sum_{m=1}\f{\ga_m}{m!}\{(z-1)\bar n(s)\}^m.
\l{10}\ee
To find consequences of this assumption let us find the mostly
probable values of $z$. The equation:
\be
n=z\f{\pa}{\pa z}\ln\Xi(z,s)
\l{11}\ee
has increasing with $n$ solutions $\bar z(n,s)$ since $\Xi(z,s)$ is
the increasing function of $z$, iff $\Xi(z,s)$ is the nonsingular at
finite $z$ function. Last condition has deep physical meaning and
practically assumes that absence of first order phase transition
\C{syw}.

Let us introduce new variable:
\be
\la=(z-1)\bar n(s).
\l{12}\ee
Corresponding eq.(\r{11}) looks as follows:
\be
\f{n}{\bar n(s)}=\le(1+\f{\la}{\bar
n(s)}\ri)\f{\pa}{\pa\la}\ln\Xi(\la).
\l{13}\ee
So, with $O(\la/\bar n(s))$ accuracy, one can assume that
\be
\la\simeq\la_c(n/\bar n(s)).
\l{14}\ee
are essential. It follows from this estimation that such scaling
dependence is rightful at least in the neighborhood of $z=1$, i.e. in
vicinity of main contributions into $\s_{tot}$. This gives:
\be
\bar n(s)\s_n(s)=\s_{tot}(s)\psi(n/\bar n(s)),
\l{15}\ee
where
\be
\psi(n/\bar n(s))\simeq\Xi(\la_c(n/\bar n(s)))\exp
\{n/\bar n(s)\la_c(n/\bar n(s))\}\leq O(e^{-n})
\l{16}\ee
is the unknown function. The asymptotic estimation follows from the
fact that $\la_c=\la_c(n/\bar n(s))$ should be, as follows from
nonsingularity of $\Xi(z)$, nondecreasing function of $n$.

The estimation (\r{14}) is rightful at least at $s\rar\infty$. The
range validity of $n$, where solution of (\r{14}) is acceptable
depends from exact form of $\Xi(z)$.  Indeed, if $\ln\Xi(z)\sim\exp
\{\ga\la(z)\}$, $\ga=const>0$, then (\r{14}) is rightful at all
values of $n$ and it is enough to have the condition $s\rar\infty$.
But if $\ln\Xi(z)\sim(1+ a\la(z))^\ga$, $\ga=const>0$, then (\r{14})
is acceptable iff $n<<\bar n^2(s)$.

Representation (\r{15}) shows that just $\bar{n}(s)$ is the natural
scale of multiplicity $n$ \C{kno}. This representation was offered
firstly as a reaction on the so called Feynman scaling for
inclusive cross section:
\be
f_k(q_1,q_2,...,q_k)\sim\prod^{k}_{i=1}\f{1}{\e(q_i)}.
\l{17}\ee

As follows from estimation (\r{16}), the limiting KNO prediction
assumes that $\s_n=O(e^{-n})$. In this regime $\X(z,s)$ should be
singular at $z=z_c(s)>1$. The normalization condition
$$\f{\pa\Xi(z,s)}{\pa z}|_{z=1}=\bar{n}(s)$$ gives:
$z_c(s)=1+\ga/\bar{n}(s)$, where $\ga>0$ is the constant.
Note, such behavior of big partition function $\X(z,s)$ is
natural for stationary Markovian processes described by logistic
equations \C{vol}.  In the field theory such equation describes the
QCD jets \C{jet}.

It is known that at $Tevatron$ energies the mean hadrons multiplicity
rise with transverse momentum. The associated mean multiplicity is

$$
C_1(q_{tr})=\bar{n}(q_{tr})=\f{\sum_n nd\s_n/d q_{tr}}{\sum_n d\s_n/d
q_{tr}}.
$$
\vskip 0.5cm
So, if
$$
C_m(q_{tr})=\ga_m\le(C_1(q_{tr})\ri)^m:~~
f_k(q_1,q_2,...,q_k)\sim\prod^{k}_{i=1}\f{1}{\e(q_i)}\O(q_{tr}),
$$
then:
$$
\bar{n}(q_{tr})\f{d\s_n/d q_{tr}}{\sum_n
d\s_n/d q_{tr}}=\Psi(n/\bar{n}(q_{tr})).
$$
This prediction is in good agreement with experiment \C{kno1}.

\section{Temperature description}\0

By definition,
\be
\s_n^{ab}(s)=\int d\o_n(q)\d(q_a+q_b-\sum^{n}_{i=1}q_i)|A_n^{ab}|^2,
\l{18}\ee
where $A_n^{ab}$ is the amplitude of $n$ creation at interaction
of particles $a$ and $b$.

Considering Fourier transform of energy-momentum conservation
$\d$-function one can introduce the generating function $\R_n$
\C{byuck}. We may find in result that $\s_n$ is defined by equality:
\be
\s_n(s)=\int^{+i\infty}_{-i\infty}\f{d\b}{2\pi}e^{\b\sqrt{s}}
\R_n(\b),
\l{3}\ee
where
\be
\R_n(\b)=\int
\le\{\prod^n_{i=1}\f{d^3q_ie^{-\b\e(q_i)}}{(2\pi)^32\e(q_i)}
\ri\}|A_n^{ab}|^2.
\l{19}\ee
The mostly probable value of $\b$ in is defined by equation of
state:
\be
\sqrt{s}=-\f{\pa}{\pa\b}\ln\R_n(\b).
\l{3'}\ee

Let us consider the simplest example of noninteracting particles:
$$
\R_n(\b)=\le\{2\pi mK_1(\b m)/\b\ri\}^n,
$$
where $K_1$ is the Bessel function. Inserting this expression into
(\r{3'}) we can find that in the nonrelativistic case ($n\simeq
n_{max}$)
$$
\b_c=\f{3}{2}\f{(n-1)}{(\sqrt{s}-nm)}.
$$
I.e., $E_{kin}=\f{3}{2}T$, where $E_{kin}=(\sqrt{s}-nm)$ is the
kinetic energy.

It is important to note that the equation(\r{3'}) have unique
real rising with $n$ and decreasing with $s$ solution $\b_c(s,n)$
\C{mart}.

The expansion of integral (\r{3}) near $\b_c(s,n)$ unavoidably gives
asymptotic series with zero convergence radii since $\R_n(\b)$ is the
essentially nonlinear function of $\b$.  From physical point of view
this means that, generally speaking, fluctuations in vicinity of
$\b_c(s,n)$ may be arbitrarily high and in this case $\b_c(s,n)$ has
not any physical sense. But if fluctuations are small (strictly
speaking, they may be arbitrarily high, bu distribution in vicinity
of $\b_c(s,n)$ should be Gaussian), then $\R_n(\b)$ should coincide
with partition function of $n$ particles and $\b_c(s,n)$ may be
interpreted as the inverse temperature.

Let us define the conditions when the fluctuations are small \C{15}.
Firstly, we should expand $\ln\R_n(\b+\b_c)$ over $\b$:
\be
\ln\R_n(\b+\b_c)=\ln\R_n(\b_c)-\sqrt{s}\b+\f{1}{2!}\b^2\f{\pa^2}
{\pa\b^2_c}
\ln\R_n(\b_c)-\f{1}{3!}\b^3\f{\pa^3}{\pa\b^3_c}\ln\R_n(\b_c)+...
\l{21}\ee
and, secondly, expend the exponent in the integral over, for
instance, over ${\pa^3\ln\R_n(\b_c)}/{\pa\b^3_c}$ neglecting higher
decomposition terms in (\r{21}). In result, $k$-th term of
the perturbation series
\be
\R_{n,k}\sim\le\{\f{{\pa^3\ln\R_n(\b_c)}/{\pa\b^3_c}}
{({\pa^2\ln\R_n(\b_c)}/{\pa\b^2_c})^{3/2}}\ri\}^k
\Ga\le(\f{3k+1}{2}\ri).
\l{22}\ee
Therefore, one should assume that
\be
{\pa^3\ln\R_n(\b_c)}/{\pa\b^3_c}<<
({\pa^2\ln\R_n(\b_c)}/{\pa\b^2_c})^{3/2}.
\l{23}\ee
to neglect this term. One of possible solution of this condition is
\be
{\pa^3\ln\R_n(\b_c)}/{\pa\b^3_c}\approx 0.
\l{24}\ee
If this condition is hold, then the fluctuations are
Gaussian, but arbitrary since theirs value is defined by
$\{{\pa^2\ln\R_n(\b_c)}/{\pa\b^2_c}\}^{1/2}$, see (\r{21}).

Let us consider now (\r{24}) carefully. We will find computing
derivatives that this condition means following approximate equality:
\be
\f{\R^{(3)}_n}{\R_n}-3\f{\R^{(2)}_n\R^{(1)}_n}{\R^2_n}+2
\f{(\R^{(1)}_n)^3}{\R^3_n}\approx 0,
\l{25}\ee
where $\R^{(k)}_n$ means the $k$-th derivative. For identical
particles (see definition (\r{19})),
\ba
\R^{(k)}_n(\b_c)=n^k(-1)^k\int \le\{\prod^n_{i=1}\e(q_i)
\f{d^3q_ie^{-\b\e(q_i)}}{(2\pi)^32\e(q_i)} \ri\}|A_n^{ab}|^2
\n\\=\s_{tot}
n^k\int \le\{\prod^k_{i=1}\e(q_i)
\f{d^3q_ie^{-\b\e(q_i)}}{(2\pi)^32\e(q_i)}
\ri\}\bar{f}_k(q_1,q_2,...,q_k),
\l{26}\ea
where $\bar{f}_k$ is the $(n-k)\geq 0$-point inclusive cross section.
It coincide with $k$-particle distribution function in the
$n$-particle system. Therefore, l.h.s. of(\r{25}) is the 3-point
correlator $K_3$:
\be
K_3\equiv\int d\o_3(q)\le(<\prod^3_{i=1}\e(q_i)>_{\b_c}-
3<\prod^2_{i=1}\e(q_i)>_{\b_c} <\e(q_3)>_{\b_c}
+2\prod^3_{i=1}<\e(q_i)>_{\b_c}\ri),
\l{27}\ee
where the index means averaging with the
Boltzmann factor $\exp\{-\b_c\e(q)\}$.

In result, to have all fluctuations in vicinity of $\b_c$ Gaussian,
we should have $K_m\approx0$, $m\geq3$. But, as follows from
(\r{23}), the set of minimal conditions looks as follows:
\be
K_m<<K_2,~m\geq3.
\l{}\ee
If experiment confirms this conditions then, independently from
number of particles, the final state may be described by one
parameter $\b_c$ with high enough accuracy $\b_c$.

Considering $\b_c$ as physical (measurable) quantity, we are forced
to assume that both the total energy of the system $\sqrt{s}=E$ and
conjugate to it variable $\b_c$ may be measured with high
accuracy\foot{Note, the uncertainty principle $\sim\hbar$ did not
restrict $\D E$ and $\D\b$.}.

\section{Real-time finite temperature generating functionals}\0

We would like to show now why and in a what conditions our $S$-matrix
interpretation of statistics is rightful.

In modern formulations, see e.g. the textbook \C{land}, the
temperature is introduced by so called periodic Kubo-Martin-Schwinger
(KMS) boundary condition \C{kms}. Namely, in the Feynman-Kac
functional integral representation of the partition function
\be
\Xi(\b)=\int D\vp e^{-S_\b(\vp)}.
\l{4.1}\ee
the action $S_\b(\vp;z)$ is defined on the Matsubara imaginary
time contour $C_M$:  $(t_i,t_i-i\b)$, but fields should obey KMS
boundary condition:
\be
\vp(t_i)=\vp(t_i-i\b).
\l{4.2}\ee
This is natural consequence of definition: $\Xi(\b)={\rm Sp}e^{-\b
{\bf H}}$.

It was offered to deform Matsubara contour in a following way:
\be
C_M\rar C_{SK}:~ (t_i,t_f)+(t_f,t_i+i\b),
\l{4.3}\ee
where $C_{SK}$ is the Mills time contour \C{mills} and $t_f>t_i$
belongs to real axis \C{pp}. Including the real-time parts
we obtain a possibility to describe time evolution of the system

But this attempt was not successful.  First of all, we have not an
evident interpretation $t_i$ and $t_f$ \C{th}.  Secondly, in spite of
real-time parts, this formulation unable to describe the time
evolution \C{11}.

\subsection{Equilibrium media}
It was shown above that if $\s_n$ is defined by (\r{18}) then one may
introduce $\R_n$ using definition (\r{19}). The Fourier transform
(\r{3}) connects $\s_n$ и $\R_n$. On other hand, $\R_n$ reminds the
partition function.

To find complete analogy with statistical physics we should consider
transition $m\rar n$ particles with amplitude $A_{nm}=<out;n|in;m>$.
Summation over $n$ and $m$ is assumed. The corresponding
$\d$-function of energy-momentum conservation law should be written
in the form:
\be
\d(\sum^n_{i=1}q_i-\sum^m_{i=1}p_i)=\int d^4P \d(P-\sum^n_{i=1}q_i)
┌\d(P-\sum^m_{i=1}p_i),~~P=(E,\vec{P}).
\l{a}\ee
This will lead to necessity introduce independently the temperature
of initial ($1/\b_i$) and final ($1/\b_f$) states. In particle
physics we can consider the final state temperature only.

In result we get to the Fourier-Mellin transform
$\R(\b,z)=\R(\b_i,z_i;\b_f,z_f)$. Direct calculations give important
factorized form:
$$
\R(\b,z)=e^{\hat{N} (\b,z;\p)}\R_0 (\p),
$$
where the operator
\ba
\hat{N} (\b,z;\p)=
\int dx dx'(\h{\phi}_+(x)D_{+-}(x-x',\b_f,z_f)\h{\phi}_-(x')-
\n\\
-\h{\p}_-(x)D_{-+}(x-x',\b_i,z_i)\h{\p}_+(x')),~\h{\p}=\f{\d}{\d\p},
\ea
acts on the functional:
\be
\R_0 (\phi_{\pm})
=\int D\Phi_+ D\Phi_-
e^{iS(\Phi_+)-iS(\Phi_-)-iV(\Phi_+ +\phi_+) + iV(\Phi_- +\phi_-)}.
\l{20}\ee
At the very end of calculations one should take auxiliary variables
$\phi_{\pm}$ equal to zero.

Here $D_{+-}$ и $D_{-+}$ are the frequency correlation functions:
$$
D_{\pm\mp}(x-x',\b)=\mp i\int d\omega(q)e^{\pm
iq(x-x'+i \mp\b)}z(q)
$$
They obey the equations:
$$
(\partial^2 +m^2)_x G_{+-}=
(\partial^2 +m^2)_x G_{-+}=0.
$$

So, all `thermodynamical' information contained in the operator
$\hat{N} (\b,z;\p)$, but interactions are described by $\R_0(\p)$.
One can say that the operator $\hat{N}$ (adiabatically) maps the
interacting filed system on the observable states. This important
property allows consider only `mechanical' processes and exclude from
consideration the `thermal' ones.

Calculating $\R_0(\p)$perturbatively one can find:
\be
\R(\b,z)=
e^{-iV(-i\hat{j}_+)+iV(-i\hat{j}_-)}
e^{ \frac{i}{2} \int dx dx'
 j_a (x)D_{ab}(x-x',\b,z)j_b(x')},
\l{19'}\ee
where $D_{++}$ is the Feynman (causal) Green function and
$$
D_{--}=(D_{++})^*
$$
is the anticausal one and, as usual, $\h{j}=\d/\d j$. At the
very end one should take $j=0$.

Let us assume now that our system is a subsystem of bigger system.
This would lead to transformation of Boltzmann factor $\exp\{-\b\e\}$
on corresponding statistics occupation number $\bar{N}(-\b\e)$. This
means that our interacting fields system is surrounded by black body
radiation. This is mechanical model of the thermostat (heat bath of
thermodynamics).

In result the matrix $D_{ab}$ takes form (we put for simplicity
$z_i=z_f=1$):
\ba
i{G} (q;\b)=
\left( \matrix{
\frac{i}{q^2 -m^2 +i\epsilon} & 0 \cr
0 & -\frac{i}{q^2 -m^2 -i\epsilon} \cr
}\right)
+\n \\ \n \\+
2\pi \delta (q^2 -m^2 )
\left( \matrix{
\tilde{n}(\frac{\b_f +\b_i}{2}|q_0 |) &
\tilde{n}(\b_i |q_0 |)a_+ (\b_i) \cr
\tilde{n}(\b_f |q_0 |)a_- (\b_f) &
\tilde{n}(\frac{\b_f +\b_i}{2}|q_0 |) \cr
}\right)
\l{31}\ea
where
$$
a_{\pm}(\b)=-e^{\frac{\b}{2}(|q_0|\pm q_0)}.
$$
Following Green functions:
$$
D_{ab}(x-x',\b)=\int \frac{d^4 q}{(2\pi)^4} e^{iq(x-x')}
{G}_{ab} (q, \b)
$$
was introduced and the occupation number
\be
n_{++}(q_0)=n_{--}(q_0)=
\le\{e^{|q_0|(\b_f +\beta_i)/2}-1\ri\}^{-1}
\equiv \tilde {n}(|q_0|\frac{\b_i +\b_f}{2}).
\l{27'} \ee
and
\be
n_{+-}(q_0)= = \Theta (q_0)(1+\tilde
{n}(q_0 \b_f))+ \Theta (-q_0)\tilde {n}(-q_0 \b_i),
\l{**}\ee
\be
n_{-+}(q_0)= \Theta (q_0)\tilde {n}(q_0 \b_i)+ \Theta
 (-q_0)(1+ \tilde {n}(-q_0 \b_f)).
\l{28} \ee

Assuming that $\b_i=\b_f =\b_c$ it is easy to find:
\be
G_{+-}(t-t')=G_{-+}(t-t'-i\beta),\;\;\;
G_{-+}(t-t')=G_{+-}(t-t'+i\beta),
\l{30}\ee
i.e. our Green function obey KMS boundary condition.

So, representation (\r{19'}) with Green functions (\r{31}) coincide
identically with (\r{4.1}) calculated perturbatively, see also
\C{pp}.

\subsection{Nonequilibrium media}

Our attempt introduce the temperature as the quantitative
characteristic of $whole$ system based on assumption that mean value
of correlators is small. We can `localize' this condition assuming
that this rough description may be extended only on subdomains of the
system. For definiteness the subdomains may be marked by space-time
coordinate $r$.

It should be underlined that we divide on the subdomains not the
system under consideration but the device, where external particles
are measured. Noting that external flow consist from noninteracting
particles (including the flow of black body radiation) the division
on subdomains can not influence on the fields interaction.

In result we introduce the `local' temperature $1/\b(r)$ for $r$-th
group of interacting particles assuming that fluctuations in vicinity
of $\b(r)$ are Gaussian. This means that the mean value of
correlation in the group is small, but the correlation between groups
may be high. Nevertheless last one is not important since the
external particles are on the mass shell. At the same time dimension
of group may be arbitrary, but large then some $r_0$ to have
possibility to introduce the temperature as the collective variable.

We can distinguish following scales. Let $L_q$ be the characteristic
4-scale of quantum fluctuations, $L_s$ be the scale thermodynamical
fluctuations and $L$ be the scale of subdomain. It is natural to
assume that $L_s>>L>>L_q$.

Corresponding generating $functional$ has the form:
$$
\R_{cp}(\alpha_1,\alpha_2)=e^{\hat{N}(\phi_a^* \phi_b)}
\R_0 (\phi_{\pm}).
$$
One may note that the `localization' gives influence on the
operator only:
$$
\hat{N}(\phi_a^* \phi_b)=
\int dY dy \hat{\phi}_a(Y+y/2) \tilde{n}_{ab}(Y,y)\hat{\phi}_b(Y-y/2),
$$
The occupation numbers $n_{ab}(Y,q)$ have same form, $\b\to\b(Y)$ and
$$
\tilde{n}_{ij}(Y,y)=\int d\omega (q) e^{iqy}n_{ij}(Y,q)
$$

We find calculating $\R_0$ perturbatively that:
\ba
\R_{cp}(\beta)=\exp\{-iV(-i\hat{j}_+)+iV(-i\hat{j}_-)\}\times
\n \\
\exp\{i\int dY dy[{j}_a (Y+y/2)G_{ab}(y,(\beta (Y)){j}_b
(Y-y/2)\} \l{4.8} \ea
where the matrix Green function $G(q,(\beta (Y)))$  was defined in
(\r{31}).

\section{Conclusion}

One more detail. Our consideration has show the
uniqueness of Bogolyubov's solution of the nonequilibrium
thermodynamics problem.  Indeed, without vanishing of correlations
perturbation series in the $\b_c$ vicinity, being asymptotic, is
divergent.

We would like to stress in conclusion that Bogolyubov's creative
works naturally unite particle and statistical physics. In result,
using Bogolyubov's mathematical basis, we have the united scientific
space in which both branches of physics, thermodynamics and quantum
field theory, supplement each other.

\vskip 0.5cm
{\bf Acknowledgement}\\
We would like to thank V.Kadyshevsky, V.A.Matveev, A.N.Tavkhelidze
for interest to described approach and E.Kuraev for fruitful
discussions.


\end{document}